# 2004 KV$_{18}$ – A visitor from the Scattered Disk to the Neptune Trojan population

Horner, J.[1] & Lykawka, P. S.[2]

[1] Department of Astrophysics, School of Physics, University of New South Wales, Sydney 2052, Australia; **j.a.horner@unsw.edu.au**
[2] Astronomy Group, Faculty of Social and Natural Sciences, Kinki University, Shinkamikosaka 228-3, Higashiosaka-shi, Osaka, 577-0813, Japan

**ABSTRACT**
We have performed a detailed dynamical study of the recently identified Neptunian Trojan 2004 KV$_{18}$, only the second object to be discovered librating around Neptune's trailing Lagrange point, L5. We find that 2004 KV$_{18}$ is moving on a highly unstable orbit, and was most likely captured from the Centaur population at some point in the last ~1 Myr, having originated in the Scattered Disk, beyond the orbit of Neptune. The instability of 2004 KV$_{18}$ is so great that many of the test particles studied leave the Neptunian Trojan cloud within just ~0.1 – 0.3 Myr, and it takes just 37 million years for half of the 91125 test particles created to study its dynamical behaviour to be removed from the Solar system entirely. Unlike the other Neptunian Trojans previously found to display dynamical instability on hundred million year timescales (2001 QR$_{322}$ and 2008 LC$_{18}$), 2004 KV$_{18}$ displays such extreme instability that it must be a temporarily captured Trojan, rather than a primordial member of the Neptunian Trojan population. As such, it offers a fascinating insight into the processes through which small bodies are transferred around the outer Solar system, and represents an exciting addition to the menagerie of the Solar system's small bodies.

# 1 INTRODUCTION

In 2001, the discovery of the first Neptunian Trojan (2001 QR$_{322}$; Chiang et al., 2003) added a new population to the zoo of Solar system small body families – the Neptune Trojans. Estimates of their number vary, but it has been suggested that the total population of Neptunian Trojans is at least as great as that of the Jovian Trojans, which are themselves thought to outnumber the population of the main asteroid belt (Chiang & Lithwick, 2005; Sheppard & Trujillo, 2006). In the first few years after the discovery of 2001 QR$_{322}$, a further five Neptunian Trojans were found – namely 2004 UP$_{10}$, 2005 TN$_{53}$, 2005 TO$_{74}$, 2006 RJ$_{103}$ and 2007 VL$_{305}$ – each of which (like 2001 QR$_{322}$) librate around Neptune's leading Lagrange point, L$_4$. The search for 'trailing' Neptunian Trojans – objects librating around Neptune's trailing Lagrange point, L$_5$, is hindered by the fact that objects around that Lagrange point are currently located in the vicinity of the constellation Scorpius in the sky, where the plane of the ecliptic passes through that of the galaxy, in essentially the same region as the galactic centre. This coincidental alignment makes it incredibly challenging to find trailing Neptunian Trojans, as they are essentially lost in the densest star fields in the whole sky[1]. Despite this difficulty, the first trailing Neptunian Trojan, 2008 LC$_{18}$, was discovered in 2008, as a result of a dedicated search program (Sheppard & Trujillo, 2010). Until recently, then, the total population of known Neptunian Trojans stood at a measly seven objects – a situation that observers have been working hard to address. Those efforts have recently borne fruit with the discovery of 2004 KV$_{18}$ – the second trailing Neptunian Trojan to be found to date[2].

A number of studies have been carried out investigating the formation and evolution of the Neptunian Trojans (Chiang et al., 2003; Chiang & Lithwick, 2005; Nesvorný & Vokrouhlicky, 2009; Lykawka et al., 2009, 2010, 2011; Lykawka & Horner, 2010). A common theme of those studies is that the Neptunian Trojans simply cannot have been formed in-situ – instead, it is thought that the Neptunian Trojans were captured during the outward migration of the giant planet after its formation, a process also invoked to explain a number of features of the various other populations of Solar system objects (e.g. Malhotra 1995; Morbidelli et al., 2005; Lykawka & Mukai, 2007a, b; Lykawka & Mukai, 2008; Minton & Malhotra, 2009; Koch & Hansen, 2011).

---

[1] The difficulties inherent in the observation of distant Solar system small bodies as they pass through this region of the sky are such that attempts to recover the Neptune Trojan 2008 LC$_{18}$ using the Australian National University's 2.3m telescope at Siding Spring Observatory, NSW, Australia in August 2011 were entirely unsuccessful – despite the fact that seeing was good and that the detection of such an object with that telescope (at an R magnitude of ~23.3) should be feasible, aside from the crowded nature of the field. For more information, we direct the interested reader to Horner et al., 2012.

[2] 2004 KV$_{18}$ was first observed on 24$^{th}$ May, 2004, from Mauna Kea, as detailed in the Minor Planets and Comets Supplement, MPS 289006; http://www.minorplanetcenter.net/iau/ECS/MPCArchive/2009/MPS_20090628.pdf. Unfortunately, despite an extensive search, the authors were unable to find out which of the telescopes atop Mauna Kea was used to make the discovery observations, and which observers should be credited with the discovery. 2004 KV$_{18}$ was fortunately discovered before it moved into the dense star fields that lie in the direction of the galactic centre. We note that the most recent observation of the object, according to the Minor Planet Centre, was obtained in May 2006, and that it has not been seen since – a consequence, no doubt, of its current journey through that densely populated region of the sky.

The idea that the Neptunian Trojans were captured, rather than having formed in-situ, helps to explain the highly excited distribution of their orbits – with eccentricities ranging as high as 0.184 (2004 $KV_{18}$) and inclinations that can exceed 25 degrees (2005 $TN_{53}$, 2007 $VL_{305}$ and 2008 $LC_{18}$). An obvious prediction of such models is that Trojans would be captured on orbits of varying dynamical stability, from those captured so strongly that they can survive as Trojans for many billions of years to those that are captured only temporarily. Such a range of capture outcomes is clearly seen in theoretical work modelling the origin of both the Neptunian and Jovian Trojan populations (e.g. Morbidelli et al., 2005; Lykawka et al., 2009, 2011; Lykawka & Horner 2010), and so it is unsurprising that recent dynamical studies have revealed that both 2001 $QR_{322}$ and 2008 $LC_{18}$ might be such dynamically unstable Neptunian Trojans (Horner & Lykawka 2010a; Horner et al., 2012, Horner & Lykawka, 2012). Indeed, recent work has revealed that one of the Jovian Trojans, (1173) Anchises, is definitively dynamically unstable (Horner, Müller & Lykawka, 2012) – adding further weight to the capture hypothesis. In this light, it is clearly important to study the dynamical behaviour of all newly discovered Neptunian Trojans, in order to see whether they fall into the dynamically stable, or dynamically unstable, camps. As more such objects are discovered, we will be able to determine what fraction of the Neptunian Trojan population are dynamically unstable, and thereby determine the contribution made by the Neptunian Trojans to the Centaur population (e.g. Horner et al., 2003, 2004a, b), and, from there, to the flux of objects colliding with the Earth (e.g. Horner & Jones, 2009, 2010, 2012; Horner & Lykawka, 2010b, c).

In this work, we present a detailed dynamical study of the recently discovered Neptunian Trojan 2004 $KV_{18}$, which features by far the highest orbital eccentricity yet observed for a Neptunian Trojan. In section 2, we detail the dynamical simulations used, before presenting our results, and a discussion of their implications, in section 3. Finally, we draw together our conclusions in section 4.

## 2 THE SIMULATIONS

In order to study the dynamical behaviour of 2004 $KV_{18}$, we followed the method used in our previous investigations of the Jovian and Neptunian Trojans (Horner & Lykawka, 2010a, 2012; Horner et al., 2012; Horner, Müller & Lykawka, 2012). We created a large population of "clones" of 2004 $KV_{18}$, spread around the nominal best-fit orbit for the object (Table 1), such that values ranging up to ±3σ from the nominal values were tested. We then followed the evolution of these clones under the gravitational influence of Jupiter, Saturn, Uranus and Neptune for a period of four billion years using the *Hybrid* integrator within the *n*-body dynamics package Mercury (Chambers, 1999). Each clone was followed until it either collided with one of the massive bodies (the Sun and the giant planets) or was ejected to a heliocentric distance of 1000 AU. This technique allows us to create dynamical maps of the stability of the object as a function of its initial orbital elements, which have been particularly useful in identifying regions of instability in previous studies of both Solar system objects and exoplanetary systems (e.g. Horner & Lykawka, 2010a; Horner et al., 2011; Robertson et al., 2012a, b). It is important to note, here, that although 2004 KV18 is the newest Neptune Trojan to be discovered, its current orbital uncertainties are far smaller than those of 2008 LC18 (the first trailing Neptunian Trojan to be identified) and are small enough to allow us to perform a detailed investigation of the dynamics of this object.

In total, we created a population of 91125 test particles, based on an orbital solution for 2004 $KV_{18}$ (Table 1) obtained on 17$^{th}$ January, 2012, from the AstDys website (http://hamilton.dm.unipi.it/astdys/). This population consisted of 25 clones in semi-major axis, *a*, 15 clones in eccentricity, *e*, 9 clones in orbital eccentricity, *i*, and 3 clones in each of the angular orbital elements, *Ω*, *ω*, and *M* (so 25x15x9x3x3x3 = 91125 test particles). As in our earlier work, the clones in a given element were spread uniformly across the full ±3σ range of plausible values[3]. The simulations were carried out on the University of New South Wales' *Katana* supercomputing cluster, and in total required somewhat more than a year of computation time to complete.

---

[3] In actuality, the errors on the various orbital elements for a given Solar system object are linked via its covariance matrix, resulting in an "error ellipse" for the object that encloses the region of element space within which the object is most likely located, to an uncertainty of ±3σ. Our simulations span a slightly larger area of orbital element space than the true 3σ error ellipse for 2004 KV18,

|   | Value | 1-σ variation | Units |
|---|---|---|---|
| a | 30.111 | 0.01094 | AU |
| e | 0.184222 | 0.001086 |  |
| i | 13.612 | 0.001472 | deg |
| Ω | 235.63 | 0.000503 | deg |
| ω | 294.329 | 0.2725 | deg |
| M | 59.951 | 0.1357 | deg |

**Table 1: The orbital elements of 2004 KV$_{18}$ at epoch MJD 56000, as obtained from the AstDys website,** http://hamilton.dm.unipi.it/astdys/ **on 17$^{th}$ January 2012.**

## 3 RESULTS AND DISCUSSION

Of the initial swarm of 91125 test particles, just 273 survived for the full four billion years of our integrations (a survival fraction of just ~0.3%), revealing that the orbit of 2004 KV$_{18}$ is incredibly unstable. Indeed, a whopping 17053 test particles were ejected from the system (or collided with one of the giant planets or the Sun) within the first ten million years of our integration. Fully half of the test particles considered (45563 objects) were ejected from the Solar system within the first 37 million years of our integrations – a level of dynamical stability more typical of a Centaur (e.g. Horner et al. 2004a, b) than a supposedly stable Neptunian Trojan. This is illustrated in Figure 1, where we present four plots showing the rate at which the number of surviving clones of 2004 KV$_{18}$ decays with time. The decay of clones of 2004 KV$_{18}$ is more than an order of magnitude more rapid than that observed for any of the unstable Trojans we have previously studied, which clearly calls into question the very nature of 2004 KV$_{18}$. Whilst it seemed perfectly feasible that the other unstable Trojans (2001 QR$_{322}$, 2008 LC$_{18}$ and (1173) Anchises) were objects that have remained in their respective Trojan clouds since the birth of the Solar system, this seems highly unlikely for 2004 KV$_{18}$. This conclusion is supported by the fact that, at the end of our simulations, not a single object remained trapped in a Neptunian Trojan-like orbit. All the survivors had semi-major axes significantly greater than that of Neptune, and only four of the 273 had perihelion distances interior to the orbit of Neptune.

In our studies of the other potentially unstable Neptune Trojans 2001 QR$_{322}$ (Horner & Lykawka, 2010a) and 2008 LC$_{18}$ (Horner et al., 2012), we found that their stability was a strong function of their initial orbital parameters, and that both objects had orbital uncertainties that spanned regions of both great dynamical stability and significant dynamical instability. As such, it is clearly interesting to consider whether the stability of 2004 KV$_{18}$ is similarly influenced by the choice of initial orbital elements used. A dynamical map showing the stability of 2004 KV$_{18}$ as a function of initial semi-major axis, *a*, and eccentricity, *e*, is presented in Figure 2.

---

since we simply take the maximal 3σ error ranges for each element (as detailed in Table 1) as the basis for our integrations. In practice, this allows us to determine whether a stable (or unstable) object lies even vaguely near a region of instability (or stability). For more details on the cloning procedure used, we direct the interested reader to our earlier works (e.g. Horner & Lykawka, 2010a).

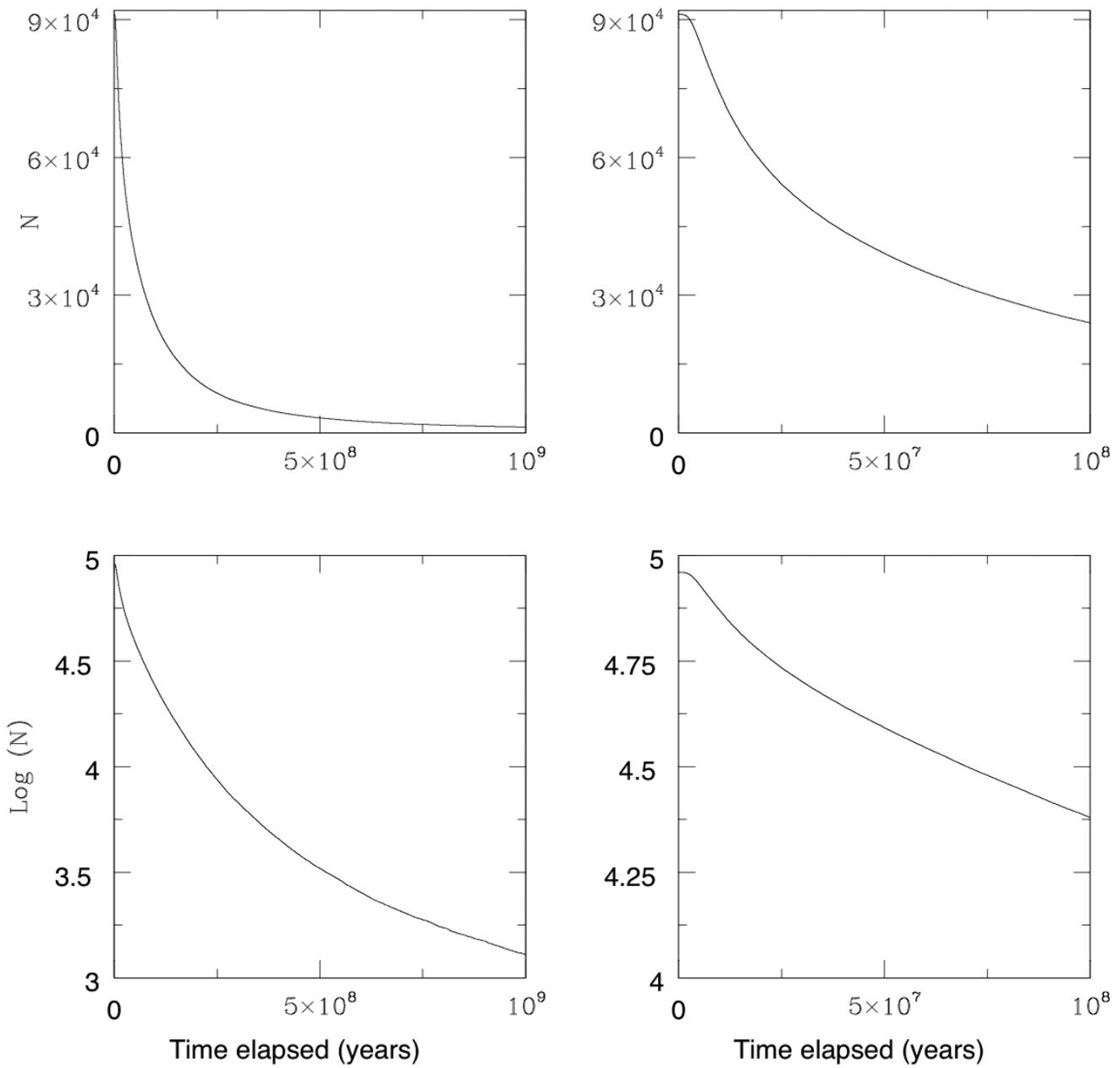

Figure 1: The rapid decay of the population of clones of 2004 KV$_{18}$ as a function of time. The left hand plots show the decay of the number of surviving clones (*N*) over the first billion years of our integrations. The right hand plots show the same information, but for the first 100 million years of our runs. The upper plots show *N* vs. time elapsed (*t*), while the lower show the decay of $\log_{10}(N)$ as a function of time. It is clear that the object exhibits extreme dynamical instability. It is also worth noting, again, that none of the survivors remained in within the 1:1 mean motion resonance with Neptune at the end of the simulations.

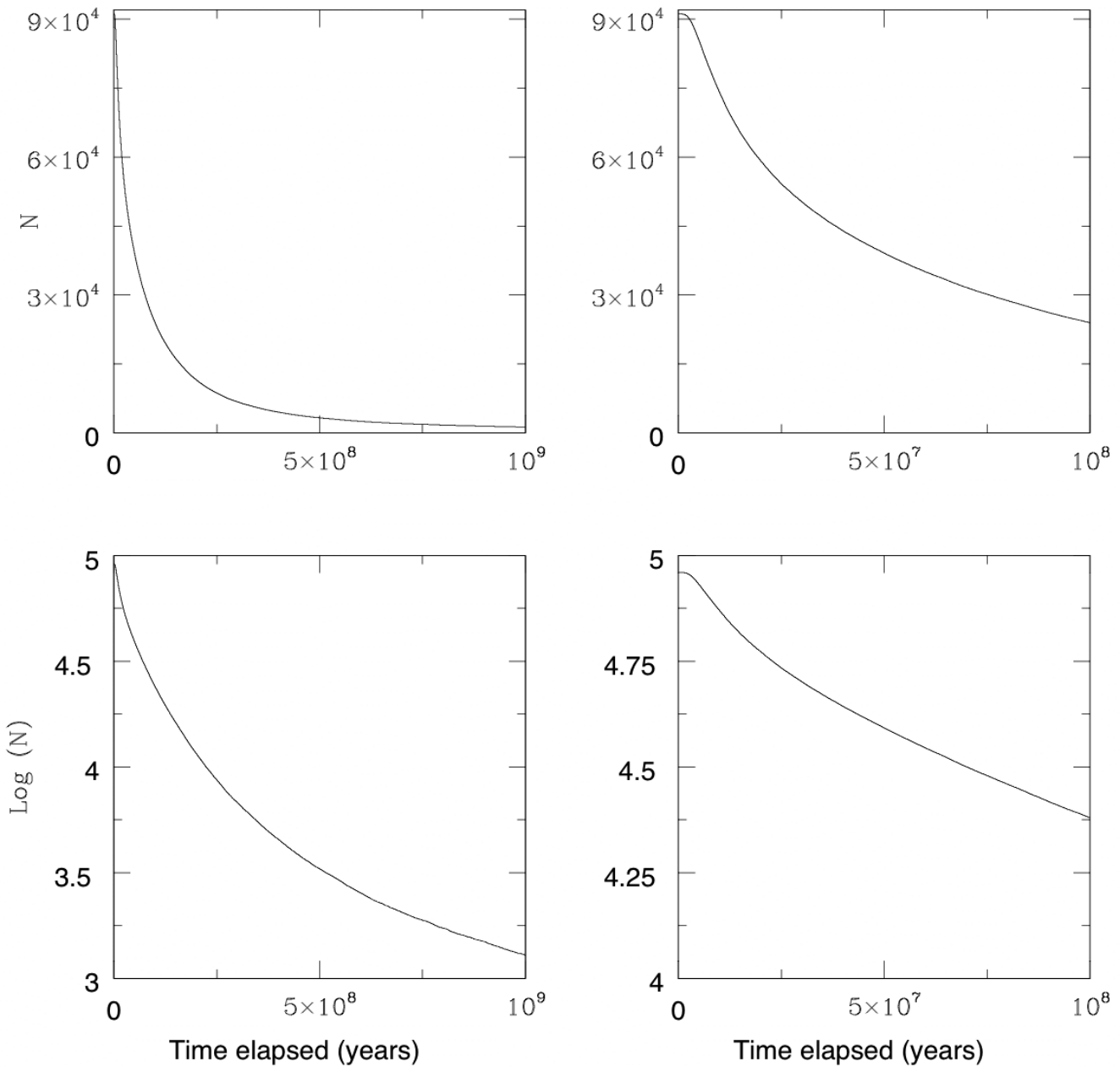

exhibit highly chaotic behaviour within the Trojan clouds on timescales of just a few Myr (as described in more detail in the main text).

It is immediately apparent that the 2004 $KV_{18}$ is strongly dynamically unstable across the whole range of allowed semi-major axes and eccentricities, much as was the case for (1173) Anchises (Horner, Müller & Lykawka, 2012). In addition, given the small orbital uncertainties of 2004 $KV_{18}$'s orbit, that uncertainty is unlikely to be the origin of the strong instability. As such, it is clear that the orbit of 2004 $KV_{18}$ is truly unstable, and that the instability is not merely the result of its orbit not yet being sufficiently well refined as to confirm that it moves in a dynamically stable region (as could be the case for both 2001 $QR_{322}$ and 2008 $LC_{18}$).

The remarkably short lifetimes displayed by clones of this object suggest that it might well be a recent and temporary capture (e.g. Horner & Evans, 2006) to Neptune's trailing Trojan cloud, rather than being an object that has resided there since the end of planetary migration. In order to test this hypothesis, we ran a small subsidiary simulation, in which we followed the evolution of 729 clones of 2004 $KV_{18}$ backwards in time for a period of 10 million years, tracking their evolution at 100-year intervals. The first (or most recent) 1.5 Myr of that simulation are shown in Figure 3. It can be seen that all clones disperse quickly, typically escaping from 1:1 mean motion resonance with Neptune (i.e., tadpole and horseshoe-type Trojan orbits) on timescales of a few hundred thousand years. At that point, the particles are no longer protected from close encounters with

Neptune, and so acquire unstable orbits marked by gravitational encounters and temporary resonant interactions with Neptune (including re-captures as Trojans) and the other giant planets. It is clear from this that the orbital evolution of 2004 KV$_{18}$ in the near past (and near future) is better represented as that of a Centaur evolving under the gravitational influence of Neptune (and, to a lesser extent, Uranus) that just happens to experience a temporary Trojan capture by Neptune around the current epoch.

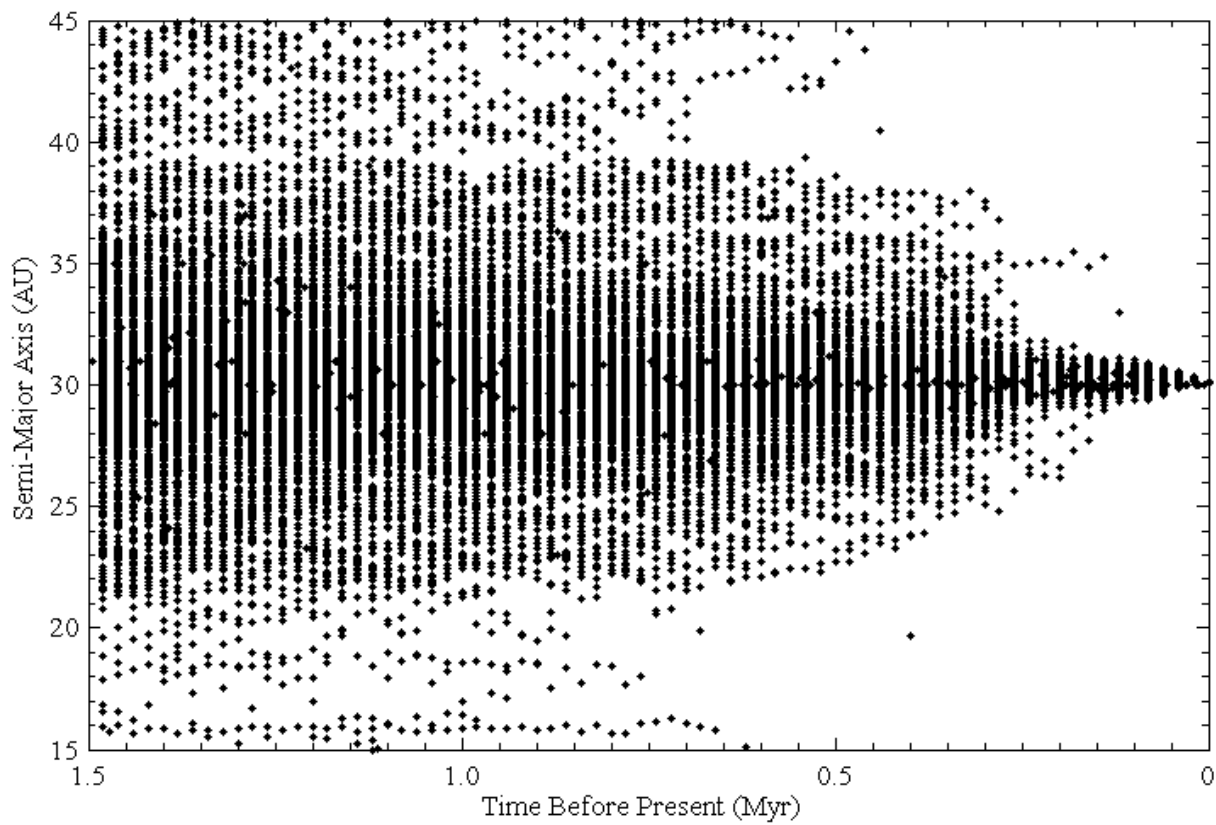

Figure 3: The evolution of the semi-major axes of 729 clones of 2004 KV$_{18}$ over the past 1.5 Myr years. For clarity, the semi-major axes are plotted at 20 kyr intervals. Note how quickly the test particles disperse from their relatively right grouping at t = 0 yr.

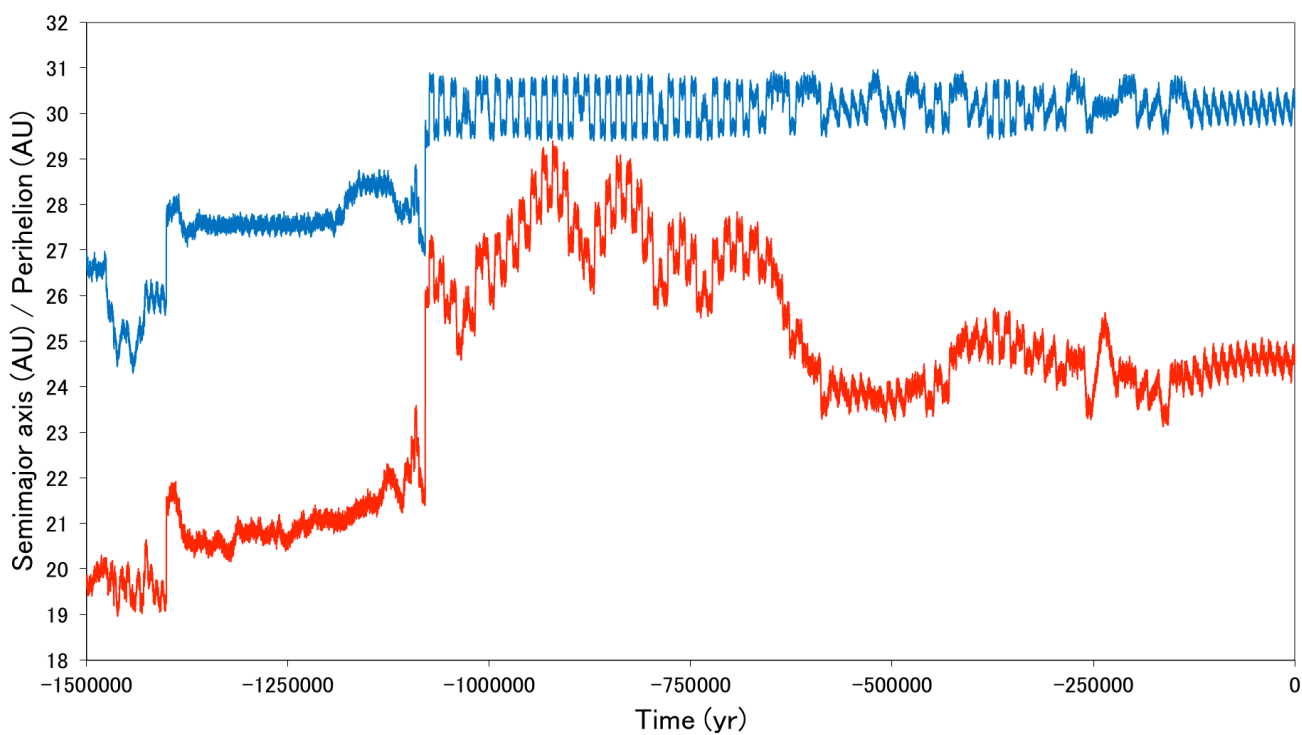

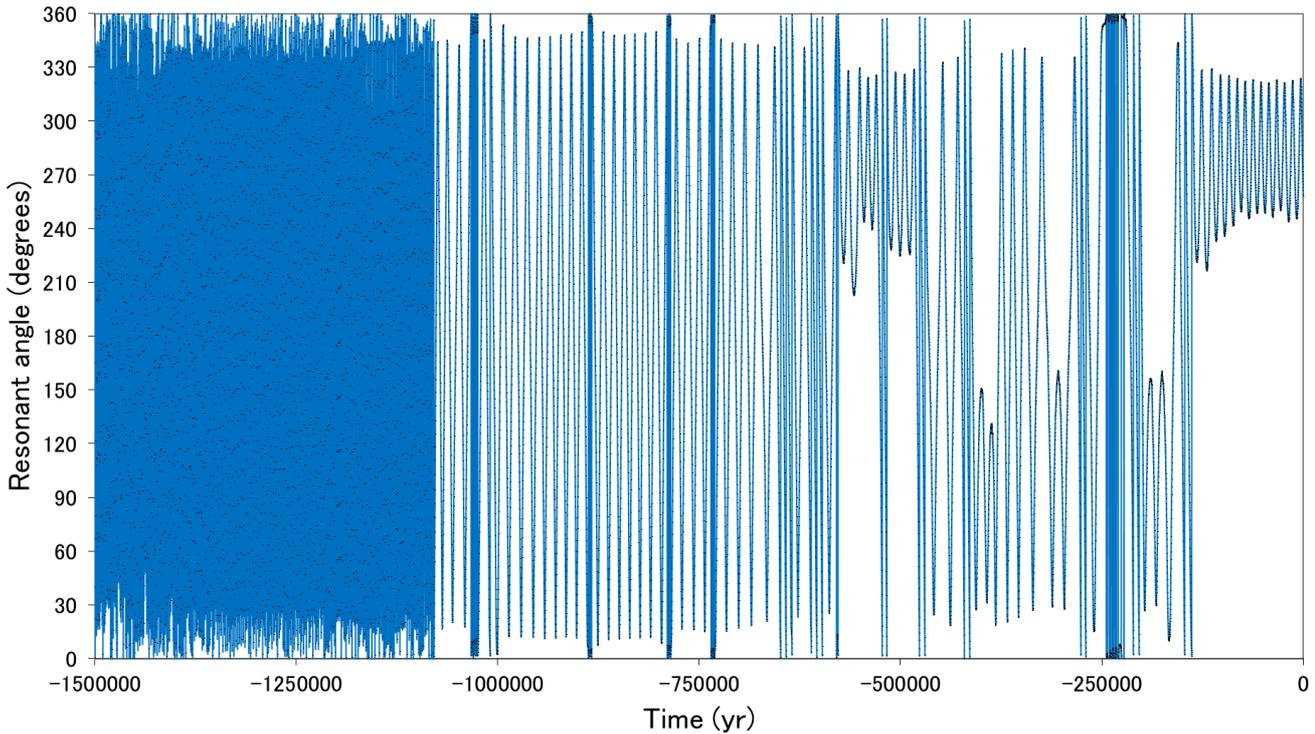

Figure 4: The evolution of the nominal best-fit orbit of 2004 KV$_{18}$ over the past 1.5 Myr. The upper panel shows the evolution of the object's semi-major axis (blue) and perihelion distance (red), as a function of time for that period. The lower panel plots the temporal evolution of the object's resonant angle, which measures the angular distance between the object and Neptune. It is immediately apparent that the object's current libration around the L$_5$ Lagrange point lasts for just ~0.13 Myr. Prior to that, the object experiences a number of temporary captures and transitions into tadpole (liberating at either L4 or L5 points) and horseshoe orbits (e.g., the period between -1.1 and -0.65 Myr). At around 1.1 Myr before the present, this object (that started on the nominal best-fit orbit of 2004 KV$_{18}$) evolves out of the Trojan region, and experiences a series of close encounters with both Neptune and Uranus.

It is also interesting to consider the instability of 2004 KV$_{18}$ in the context of libration in the resonant motion, as represented by the resonant angles and libration amplitudes, $A$[4], displayed by its various clones. Figure 4 shows a representative case of the typical orbital history experienced by clones of 2004 KV$_{18}$ in our backwards integrations. The object in question (which began on the nominal best-fit orbit for 2004 KV$_{18}$) spends only ~0.13 Myr in its current trailing tadpole orbit (L5), and prior to that period displays a remarkably chaotic evolution, with temporary captures/transitions into tadpole (both L4 and L5) and horseshoe orbits. In this case, orbital evolution within the Trojan cloud is confined to just the last 1.1 Myr, although that particular clone did go on to exhibit further short periods of Trojan-like behaviour as a result of additional temporary captures in Neptune's 1:1 mean-motion resonance.

When we examine the behaviour of the whole sample of 729 test particles considered in our backward integrations (as shown in Fig. 3), we find that they all exhibit similar behaviour to that shown in Fig. 4. Typically, clones escape from their tadpole-orbits around the L$_5$ Lagrange point on timescales between 0.1 and 0.3 Myr. Following their escape, they display strong chaotic orbital evolution, which causes repeated transitions between tadpole (L4/L5) and horseshoe orbits for a period of several hundred thousand years. During that period, an ever increasing number escape entirely from the 1:1 mean-motion resonance. Overall, the clones also tend to leave the Trojan cloud (considering both tadpole and horseshoe orbits) on timescales of 1±0.5 Myr.

---

[4] As discussed in earlier work, the libration amplitude, $A$, details the scale of the full angular motion of a given Trojan during its libration around its host Lagrange point, and appears to be inextricably linked to the dynamical stability (or lack of stability) of other Trojan objects (e.g. Horner & Lykawka, 2010a; Horner et al., 2012).

Earlier works have established that the overall (in)stability of a given Trojan can generally be estimated simply through the determination of that Trojan's libration amplitude. Those studies reveal that, typically, objects with $A > 50$-$60$ deg are likely to become unstable over timescales shorter than the age of the solar system (Nesvorny & Dones 2002; Marzari et al., 2003; Zhou et al., 2009). We estimated the libration amplitudes for three representative clones of 2004 $KV_{18}$ which spanned the full range of allowed orbits. The first clone considered had initial orbital elements that were all three sigma smaller than the nominal best-fit values. The second was the nominal best-fit orbit, and the third had initial elements that were all three sigma greater than the best-fit values. The values of $A$ for these clones were approximately 68, 77, and 102 deg, with uncertainties of ~2-3 deg (See also Figure 4). It is clear that these values are larger than the 50-60 deg threshold that delimits long-term stability within the Trojan cloud, as detailed above. This therefore agrees with the idea that 2004 $KV_{18}$ is currently evolving on a highly unstable orbit as a result of a recent capture into the L5 Trojan swarm in the recent past. In addition, the fact that several clones of 2004 $KV_{18}$ experience temporary captures to both the L5 and L4 Trojan populations, over the 10 Myr simulated, on orbits that feature libration amplitudes comparable to those determined above adds yet further weight to the idea that 2004 $KV_{18}$ is a recently captured Trojan rather than a member of a primordial decaying population of Neptune Trojans.

The fact that 2004 $KV_{18}$ becomes a Centaur (i.e. a dynamically unstable giant planet-crossing/approaching object) upon leaving the Neptunian Trojan cloud confirms the existence of a steady population of dynamically fresh Centaurs[5] interacting with Neptune. Given that the dynamical evolution of small bodies in the Solar system is a time-reversible process (an assumption reasonably valid over the last 4 Gyr), and recalling that the dynamical lifetimes of Centaurs are typically several or a few tens of Myr, it follows that any of our test particles that survive for the full 4 Gyr have most likely evolved into one of the reservoirs that currently feed such fresh Centaurs in the "recent" Solar system. In line with this idea, the most significant candidate reservoir is thought to be the Scattered Disk, within which many TNOs on Neptune-encountering orbits have been identified (e.g., Lykawka and Mukai 2007a; Gladman et al. 2008). Figure 5 illustrates the final orbits occupied by the surviving clones of 2004 $KV_{18}$ after 4 Gyr, and can be considered as being representative of the unstable objects sourcing the Centaur-like objects required to explain 2004 $KV_{18}$'s current orbit.

Figure 5 reveals that the surviving clones of 2004 $KV_{18}$ acquire a variety of orbits that strongly resemble those of TNOs evolving in the Scattered Disk (with perihelion distances beyond the orbit of Neptune, and large orbital eccentricities and inclinations). Even more interesting, the surviving clones include objects that have become "detached" from the gravitational domain of Neptune as a result of temporary captures in mean motion resonances (See Lykawka and Mukai 2007b for more details). These results suggest that 2004 $KV_{18}$ is most likely a former Scattered Disk object that has become temporarily captured as a Neptune Trojan as part of its transition to (or from) the Centaur population, in much the same way as other such objects experience temporary captures in other mean motion resonances beyond Neptune.

---

[5] Objects on unstable orbits located beyond Neptune that recently acquired a Centaur-like orbit (i.e., $q < 30.1$ AU).

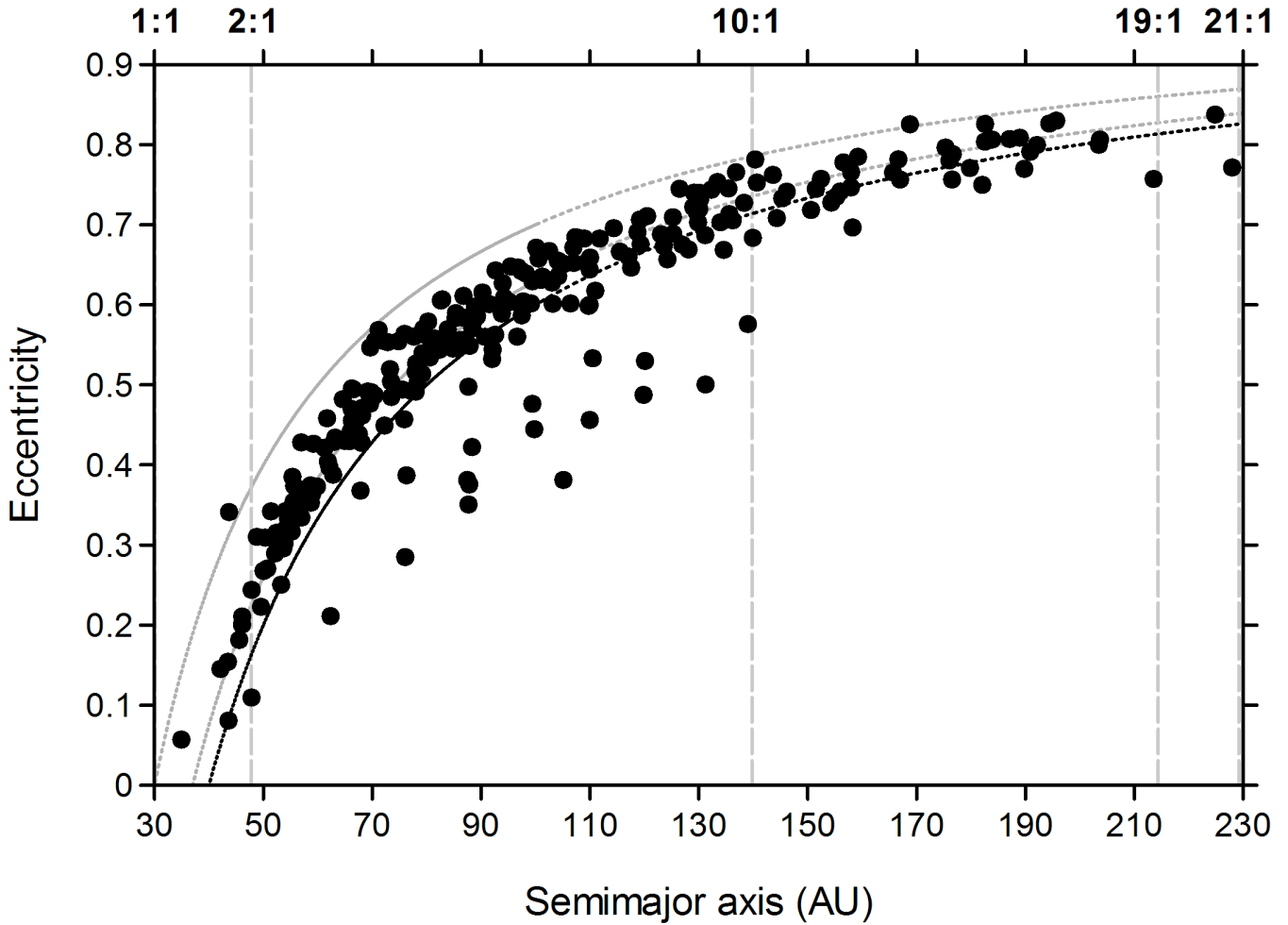

Figure 5: The final orbital elements of the 273 clones of 2004 KV$_{18}$ (from an initial population of 91125) that survived the full 4 Gyr of dynamical evolution in our simulations. Dotted curves denote lines of constant perihelion distance located at $q = 30$, 37 and 40 AU (from left to right). Vertical dashed lines show the locations of several interesting mean motion resonances with Neptune. The majority of objects that have evolved to orbits with $q > 40$ AU are locked in mean motion resonances described by $r$:1 ratios, with semi-major axes ranging as high as ~230 AU.

## 4 CONCLUSIONS

We have performed a highly detailed analysis of the dynamical evolution of the newly discovered Neptune Trojan 2004 KV$_{18}$. Our results show that this unusual object is too dynamically unstable to be considered a native of the Neptunian Trojan population, and instead imply that it is most likely a relatively recent capture to the Neptunian Trojan cloud. Given the remarkably rapid decay in the population of clones of 2004 KV$_{18}$ (with the majority of clones leaving the Trojan cloud in just a few Myr), it is highly unlikely that 2004 KV$_{18}$ is a representative of a once much larger population of primordial Neptunian Trojans on similarly unstable orbits. If one were instead to assume that 2004 KV$_{18}$ *was* a relic survivor of a once larger slowly decaying primordial Trojan population, that scenario would simply require the primordial presence of many orders of magnitude more mass than any plausible values that have been determined through models of Neptune Trojan formation (e.g. Lykawka et al. 2009, 2011; Nesvorny and Vokrouhlicky 2009). In sum, the highly chaotic orbital behaviour of representative orbits of 2004 KV$_{18}$ is characteristic of a "recently" (temporarily) captured object – a conclusion supported by the fact that none of the 91125 test particles considered in our main integration survived as a Neptune Trojan at the end of our integrations.

In agreement with the unstable orbital nature of 2004 KV$_{18}$, we found that the libration amplitudes of test particles spanning the three sigma error range of its allowed orbits lie within ~70-100 deg, which are values clearly larger than those (50-60 deg) that delimit the region of stability in the Trojan cloud. It is therefore no surprise that the object exhibits a highly chaotic evolution during its typically short lifetime as a Neptune

Trojan, and that its clones typically escape from the Neptunian Trojan population on very short timescales. Indeed, the current libration of 2004 KV$_{18}$ about Neptune's L5 Lagrange point has likely only been occurring for the past 0.1 – 0.3 Myr!

Whilst it is not possible to definitively constrain the origin of 2004 KV$_{18}$'s current Trojan orbit, the simplest explanation is this object represents a dynamically fresh Centaur that only relatively recently escaped the Scattered Disk population. As part of its chaotic evolution in the Outer Solar system, it has only "recently" been captured to the Neptunian Trojan cloud, and will escape again in the astronomically near future. Such dynamically-driven captures have been shown to be feasible for the giant planets (e.g. Horner & Evans, 2006). We note, in passing, that such a temporary capture hypothesis has invoked for the origin of the recently discovered Earth Trojan, 2010 TK$_7$ (Connors et al., 2011; Dvorak et al., 2012).

Unlike the other Neptunian Trojans that have been shown to display significant dynamical instability (2001 QR322 – Horner & Lykawka, 2010a; 2008 LC18 – Horner et al., 2012), we find that the instability displayed by 2004 KV$_{18}$ is independent of the initial orbital elements chosen for the object. Such behaviour is, superficially, similar to that displayed by the Jovian Trojan (1173) Anchises, which was recently shown to be dynamically unstable on timescales of several hundred million years (Horner, Müller & Lykawka, 2012). In both cases, the lack of variation in stability as a function of the initial orbital elements used is a direct result of the precision with which the orbits are known. In this case, that lack of variation acts to strengthen the case for 2004 KV$_{18}$ being a recent capture to the Neptunian Trojan population, rather than a primordial member of that population[6].

Over the coming years, it is likely that the known Neptunian Trojan population will grow significantly as the next generation of all-sky surveys (such as Pan-STARRS and LSST) start to yield results. Once we can be reasonably confident that the known population of Neptunian Trojans is complete to a given size regime, then it should be possible to use dynamical studies of those Trojans to estimate the steady-state populations of dynamically fresh Centaurs and Scattered Disk objects that are currently strongly interacting with Neptune (the "scattering" objects described in Gladman et al. 2008). The detection of more objects like 2004 KV$_{18}$ will then allow the populations of those unstable reservoirs to be calculated, allowing us to obtain firm estimates of the frequency, and duration, with which such objects can be expected to be captured as Neptunian Trojans.

**ACKNOWLEDGEMENTS**
The authors wish to thank the referee, E. Chiang, for providing a swift, positive, and helpful review of our work. JH acknowledges the financial support of the Australian Research Council, through the ARC discovery grant DP774000. They simulations carried out in this work were performed using the *n*-body dynamics package *MERCURY* (Chambers, 1999), and were based on orbits obtained from the *AstDys* website (http://hamilton.dm.unipi.it/astdys/), which is maintained by funding from the University of Pisa, the Agenzia Speziale Italiana, the Astronomical Observatory of Belgrade and the University of Valladolid, and uses observations published by the IAU Minor Planet Centre (http://www.minorplanetcenter.net/iau/mpc.html).

---

[6] During the revision of this work, the referee brought a recent posting on the arXiv server by Guan, Zhou and Li to our attention (namely Guan, Zhou & Li, 2012). Those authors have independently studied the dynamics of 2004 KV$_{18}$, and also conclude that it is most likely a recently captured object, rather than one native to the Neptunian Trojan cloud.